\documentclass[twocolumn,prl,epsfig,amsmath,amssymb]{revtex4}
\usepackage{graphicx}
\usepackage{color}
\usepackage{amssymb}
\usepackage{latexsym, amsmath}
\usepackage{pst-plot, pstricks}

\usepackage[arrow, matrix, curve]{xy}

\newcommand{\Tr}{{\rm Tr}}
\newcommand{\hr}{{\mathcal{H}}}
\newcommand{\C}{{\mathbb C}}
\newcommand{\R}{{\mathbb R}}
\newcommand{\N}{{\mathbb N}}
\newcommand{\idn}{\mathbf{1}}
\newcommand{\proof}{{\bf Proof.} }
\newtheorem{theorem}{Theorem}
\newtheorem{lemma}[theorem]{Lemma}

\newtheorem{definition}[theorem]{Definition}

\newtheorem{conjecture}[theorem]{Conjecture}
\newtheorem{example}[theorem]{Example}
\newcommand{\nix}{{\rule{0pt}{2pt}}}
\newcommand{\qedd}{{\nix\nolinebreak\hfill\hfill\nolinebreak$\Box$}}
\newcommand{\qed}{{\qedd\par\medskip\noindent}}
\newcommand{\lineclear}{{\rule{0pt}{0pt}\nopagebreak\par\nopagebreak\noindent}}

\begin{document}

\title{Convex Trace Functions on Quantum Channels and the Additivity Conjecture}

\author{Markus M\"uller\footnote{E-mail: mueller@math.tu-berlin.de}\\}
\mbox{}\vskip 0.5cm \affiliation{
$^1$Institut f\"ur Mathematik, Technische Universit\"at Berlin, Stra\ss e des 17.~Juni 136, 10623 Berlin, Germany
\\
$^2$Max Planck Institute for Mathematics in the Sciences, Inselstr.~22, 04103 Leipzig, Germany\\
}
\date{May 25, 2009}
\begin{abstract}
We study a natural generalization of the additivity problem in quantum information theory: given a pair of quantum channels,
then what is the set of convex trace functions that attain their maximum on unentangled inputs, if they are applied to the corresponding output
state?

We prove several
results on the structure of the set of those convex functions that are ``additive'' in this more general sense. In particular, we show that
all operator convex functions are additive for the Werner-Holevo channel in $3\times 3$ dimensions, which contains the well-known additivity
results for this channel as special cases.
\end{abstract}
\maketitle

\setcounter{secnumdepth}{1}

\section{Introduction and Main Definition}
For quite some time, the additivity conjecture has been one of the most notorious open problems in
quantum information theory; it has been settled only recently in a breakthrough paper by Hastings~\cite{Hastings}.
The original conjecture can be stated in several equivalent ways~\cite{Shor}; one possible formulation is via
the {\em minimum output entropy} of a quantum channel $\Phi$, defined as
\[
   S^{min}(\Phi):=\min_\rho S(\Phi(\rho))=\min_\rho \Tr \left(\strut -\Phi(\rho)\log\Phi(\rho)\right),
\]
where the minimization is over all input states $\rho$, and $S$ is von Neumann entropy.
The intuition is that $S^{min}(\Phi)$ is a measure of noisiness
of the channel $\Phi$.

The original additivity conjecture stated that
\begin{equation}
   S^{min}(\Phi\otimes\Omega)=S^{min}(\Phi)+S^{min}(\Omega)
   \label{AddConj}
\end{equation}
for all channels $\Phi$ and $\Omega$; that is, the minimum output entropy of a pair of channels should be the
sum of the individual minimum output entropies. During the years that the problem has been studied, it turned out to be convenient
to generalize the additivity problem to {\em $p$-R\'enyi entropies}: For $p>0$, $p\neq 1$, and density matrices $\rho$, define~\cite{MinRank}
\[
   S_p^{min}(\Phi):=\min_\rho \frac 1 {1-p} \log \Tr \left(\strut\Phi(\rho)^p\right),
\]
and then the question is whether
\begin{equation}
   S_p^{min}(\Phi\otimes\Omega)=S_p^{min}(\Phi)+S_p^{min}(\Omega)
   \label{AddConjp}
\end{equation}
holds true in general, for all $p>0$. Due to the limit $S_1^{min}(\Phi):=\lim_{p\to 1} S_p^{min}(\Phi)=S^{min}(\Phi)$,
(\ref{AddConjp}) is a natural generalization of~(\ref{AddConj}).

Quite surprisingly, it turned out that the conjectured equalities~(\ref{AddConjp}) and~(\ref{AddConj}) are both false
in general. They were subsequently disproved by constructing counterexample channels, first for $p>4.79$~\cite{HolevoWerner},
then for $p>2$~\cite{Winter}, then for $p>1$ and $p\approx 0$~\cite{Hayden,MinRank,WinterComm}, and
finally for $p=1$~\cite{Hastings}, killing the original conjecture~(\ref{AddConj}).
For detailed expositions of the problem and its history, see for example \cite{WinterComm} or \cite{HolevoExposition}.

Despite those no-go results, it has been shown that additivity holds for many interesting classes of channels
and several values of $p$,
for example for the cases that one of the channels is the identity channel~\cite{AHW,Amosov} or a unital qubit channel~\cite{King}.
Even if additivity fails in general, its validity in special cases is still interesting in its own and potentially
useful for channel coding problems, cf.~\cite{KoenigWehner}. The main goal of this paper is to show that some of those
results for special channels have a natural interpretation within a more general framework.

To motivate our more general definition, notice first that the additivity conjecture 
can equivalently be stated as the assertion that {\em entanglement does not help} to produce pure outputs.
In fact, Equation~(\ref{AddConj}) holds if and only if the map
\[
   \rho\mapsto S\left(\strut\Phi\otimes\Omega(\rho)\right)
\]
attains its global minimum at an {\em unentangled} input state $\rho$:
since $S(\sigma\otimes\rho)=S(\sigma)+S(\rho)$ for density operators $\sigma$ and $\rho$, we get
\[
   S^{min}(\Phi\otimes\Omega)\leq S(\Phi\otimes\Omega(\rho_\Phi\otimes\rho_\Omega))=S^{min}(\Phi)+S^{min}(\Omega)
\]
if $\rho_\Phi$ and $\rho_\Omega$ are the minimizers for the two channels, i.e. $S^{min}(\Phi)=S(\Phi(\rho_\Phi))$
and similarly for $\rho_\Omega$. This means that the inequality ``$\leq$'' in~(\ref{AddConj}) is always true.

On the other hand, as von Neumann entropy is concave, the global minimum will be attained at some
{\em pure} input state; also, $\rho_\Phi$ and $\rho_\Omega$ can be chosen pure.
Thus, the fact that $\rho_\Phi\otimes\rho_\Omega$ is indeed the global minimizer, i.e. ``$=$'' holds
in (\ref{AddConj}), is equivalent to the fact that no other entangled input state can produce even smaller output
entropy.

Equation~(\ref{AddConjp}) can be reformulated in a similar way: additivity for $p$-R\'enyi entropy with $p>1$
holds true if and only if the function 
\[
   \rho\mapsto \Tr \left(\Phi\otimes\Omega(\rho)^p\right)
\]
attains its global maximum at an unentangled input state $\rho$.
Thus, we have two variations of the same general problem:
{\em Compute the trace of a convex function of the output, and decide whether this expression attains its global
maximum at an unentangled input state.} For von Neumann entropy $S$, this function is $x\log x$, while
for the $p$-R\'enyi entropy $S_p$ with $p>1$, this function is $x^p$.

It is natural to ask what happens if the problem is generalized. What if one takes another convex function,
different from $x\log x$ or $x^p$?
We use Definition~\ref{DefEntFn} to study the generalized problem.
In this definition and all of the following, applying a (convex) function $f:[0,1]\to\R$ to a density matrix $\sigma:=\Phi\otimes\Omega(\rho)$
is meant in the sense of spectral calculus: diagonalizing $\sigma=U{\rm diag}(\lambda_1,\ldots,\lambda_n)U^\dagger$,
we define
\[
   f(\sigma):=U\left(\begin{array}{ccc}f(\lambda_1) & & \\ & \ddots & \\ & & f(\lambda_n)\end{array}\right)U^\dagger
\]
such that in particular $\Tr f(\sigma)=\sum_{i=1}^n f(\lambda_i)$, where the sum is over all eigenvalues $\lambda_i$ of $\sigma$.

\begin{definition}[Additive Functions on Q-Channels]
\label{DefEntFn}
Let $f:[0,1]\to\R$ be a convex function, and let $\Phi$ and $\Omega$ be
quantum channels. We say that $f$ is {\em additive} for $(\Phi,\Omega)$ if
there exists some unentangled input state $\rho_u$ such that
\begin{equation}
   \Tr f(\Phi\otimes\Omega(\rho_u))\geq\Tr f(\Phi\otimes\Omega(\sigma))
   \label{EqDef}
\end{equation}
for all input states $\sigma$.
\end{definition}
There are some simple consequences of this definition. First note that if $f$ is convex
as a real function, then it is automatically a ``convex trace function'' on the density operators
in the sense that
\[
   \Tr f(\lambda\rho+(1-\lambda)\sigma)\leq \lambda\Tr f(\rho)+(1-\lambda)\Tr f(\sigma),
\]
see \cite{Lieb}. Thus, $\Tr f(\cdot)$ attains its maximum on pure input states, i.e. (\ref{EqDef})
is equivalent to the existence of pure states $\psi_1$ and $\psi_2$ such that
\[
   \Tr f(\Phi(\psi_1)\otimes \Omega(\psi_2))\geq \Tr f(\Phi\otimes\Omega(\varphi))\quad\forall \mbox{ pure states }\varphi.
\]
Yet, in contrast to von Neumann or $p$-R\'enyi entropy, there
is in general no way to further simplify the expression on the left-hand side by splitting the function of
the tensor product into two addends or factors.

Clearly, this definition captures the additivity problems as special cases:
\begin{lemma}
\label{Lemma2}
Let $(\Phi,\Omega)$ be a pair of quantum channels. Then additivity of $p$-R\'enyi entropy
\[
   S_p^{min}(\Phi\otimes\Omega)=S_p^{min}(\Phi)+S_p^{min}(\Omega)
\]
holds if and only if the function $f_p$ is additive for $(\Phi,\Omega)$, where
\[
   f_p(x):=\left\{
      \begin{array}{cl}
         x^p & \mbox{if }p>1,\\
         x\log x & \mbox{if }p=1, \\
         -x^p & \mbox{if }0<p<1.
      \end{array}
   \right.
\]
\end{lemma}
Hence proving the additivity conjecture for a pair of channels is equivalent to showing that $x\log x$ is additive.
Is there any reason why this more general approach could help?
In fact, there is a popular example in matrix analysis where a similar strategy turned out
to be successful, which is L\"owner's theory of operator convex functions (\cite{Bhatia,Nielsen}).

A real function $f$ is called {\em operator convex} if
\begin{equation}
   f(\lambda\rho+(1-\lambda)\sigma)\leq \lambda f(\rho)+(1-\lambda)f(\sigma)
   \label{EqOperatorConvex}
\end{equation}
for all self-adjoint operators $\rho$ and $\sigma$ and $0<\lambda<1$. This is an operator inequality, meaning that
the difference of the right- and left-hand side is positive semidefinite. Clearly, operator
convex functions are convex, but the converse turns out to be false. For example, $x^3$ is convex,
but not operator convex.

Given some convex function $f$, it can be difficult to decide directly from the definition (\ref{EqOperatorConvex})
whether $f$ is operator convex. By contrast, it turns out that there is a simple characterization of
the set of {\em all} operator convex functions, which can be stated elegantly in terms of integral representations
or complex analysis.
This is an unexpected result, since the definition (\ref{EqOperatorConvex}) itself
involves only a linear-algebraic inequality.

Thus, it seems reasonable to hope that something similar might happen in the case of the additivity
problem, at least for special classes of (highly symmetric) channels:
possibly the class of additive functions for a channel pair, defined by the linear-algebraic inequality~(\ref{EqDef}),
is also simple to characterize. As we will show in this paper, this speculation turns out to be true
for the Werner-Holevo channel in $3\times 3$ dimensions at least.

We start by giving some simple examples.

\section{Some Examples}
\begin{example}
\label{LemAffine}
Let $a>0$ and $b,c\in\R$.
A convex function $f:[0,1]\to\R$ is additive for a pair of channels if and only if the
function
\[
   a f(x)+bx+c
\]
is additive for that pair of channels.
In particular, linear functions $f(x)=bx+c$
are additive (for every pair of quantum channels).
\end{example}
\proof It is clear that scaling a function with $a>0$ does not change the location of its global maximum.
If $\Phi:\mathcal{S}(\hr_1^\Phi)\to\mathcal{S}(\hr_2^\Phi)$ and $\Omega:
\mathcal{S}(\hr_1^\Omega)\to\mathcal{S}(\hr_2^\Omega)$ are arbitrary quantum channels, and if $\rho$
is an arbitrary state on $\hr_1^\Phi\otimes\hr_1^\Omega$, then it holds for $f(x):=bx+c$
\begin{eqnarray*}
\Tr f(\Phi\otimes\Omega(\rho))&=&\Tr(b\cdot \Phi\otimes\Omega(\rho)+c\cdot \idn)\\
&=&b+c\cdot \dim \hr_2^\Phi\cdot \dim \hr_2^\Omega.
\end{eqnarray*}
Thus, $\Tr f(\Phi\otimes\Omega(\rho))$ is constant, and can be added to any function
without changing its additivity properties. In particular, $f$ itself is additive,
as every unentangled input $\rho_u$ satisfies (\ref{EqDef}).
\qed

\begin{example}
\label{ExIdentity}
For channels of the form $\Phi\otimes\idn$, every convex function is additive.

That is,
if $\Phi$ is an arbitrary quantum channel, and $\idn$ is the identity channel on some Hilbert space,
then every convex function $f:[0,1]\to\R$ is additive for $(\Phi,\idn)$.
\end{example}
In particular, as is well-known, von Neumann entropy and the $p$-R\'enyi entropies are additive for such channels for all $p>0$.

\proof
The proof closely follows the lines of \cite{HolevoWerner}. Suppose $\rho_{12}'=(\Phi\otimes\idn)(\rho_{12})$.
We may choose $\rho_{12}$ to be pure. Let $U_{13}$ be a unitary dilation of $\Phi$, such that
\[
   \rho_{12}'=\Tr_3(U_{13}\otimes\idn_2)(\rho_{12}\otimes|\varphi_3\rangle\langle\varphi_3|)(U_{13}^\dagger\otimes \idn_2).
\]
As seen in Example~\ref{LemAffine}, we may assume that $f(0)=0$.
Since the expression right of $\Tr_3$ is a pure state, the spectrum will not change if we replace $\Tr_3$ by
$\Tr_{12}$ up to possible multiplicity of the eigenvalue zero. Thus,
\begin{eqnarray*}
   \Tr f(\rho_{12}')&=&\Tr f(\Tr_{12}
   (U_{13}\otimes\idn_2)(\rho_{12}\otimes|\varphi_3\rangle\langle\varphi_3|)(U_{13}^\dagger\otimes \idn_2))\\
   &=&\Tr f(\Tr_1 U_{13}(\rho_1\otimes|\varphi_3\rangle\langle\varphi_3|)U_{13}^\dagger),
\end{eqnarray*}
where $\rho_1:=\Tr_2\rho_{12}$. By convexity, this expression is maximized if $\rho_1$ is pure, i.e.
$\rho_{12}=\rho_1\otimes\rho_2$.\qed

It is well-known~\cite{Hayden} that the minimum output $p$-R\'enyi entropy of a pair of quantum channels is at least
as large as that of one of its constituents, i.e. $S_p^{min}(\Phi\otimes\Omega)\geq S_p^{min}(\Phi)$.
The following lemma generalizes this statement, and yields an analogous property for all convex functions.
\begin{lemma}[Single Channel Bound]
\label{LemSingleChannel}
\lineclear
If $f:[0,1]\to\R$ is a convex function with $f(0)=0$, then
\[
   \max_\rho \Tr f(\Phi\otimes\Omega(\rho))\leq \max_\rho \Tr f(\Phi(\rho))
\]
and similarly for $\Omega$.
\end{lemma}
{\bf Remark.} If $f(0)\neq 0$, then the bound is $\max_\rho \Tr f(\Phi(\rho))+d_\Phi(d_\Omega-1)f(0)$, where
$d_\Phi$ and $d_\Omega$ denote the dimensions of the output Hilbert spaces of $\Phi$ and $\Omega$ respectively.

\proof
If $|0\rangle$ denotes an arbitrary pure state on the output Hilbert space of $\Omega$,
and $\{\lambda_i\}_{i=1}^{d_\Phi}$ is the spectrum of $\Phi(\rho)$, then the spectrum of
$\Phi(\rho)\otimes|0\rangle\langle 0|$ is $\{\lambda_1,\ldots,\lambda_{d_\Phi},0,0,\ldots,0\}$,
with $d_\Phi d_\Omega-d_\Phi$ zeroes. Thus,
\begin{eqnarray*}
   \max_\rho \Tr f(\Phi\otimes\Omega(\rho)) &=& \max_\rho \Tr f(\Phi\otimes\idn(\idn\otimes\Omega(\rho)))\\
   &=&\max_{\rho'=\idn\otimes\Omega(\rho)} \Tr f(\Phi\otimes\idn(\rho'))\\
   &\leq& \max_{\rho'} \Tr f(\Phi\otimes\idn(\rho'))\\
   &\stackrel {(*)} =& \max_{\rho'\mbox{ unentangled}} \Tr f(\Phi\otimes\idn(\rho'))\\
   &=&\max_{\rho_A,\rho_B} \Tr f(\Phi\otimes\idn(\rho_A\otimes\rho_B))\\
   &=&\max_\rho \Tr f(\Phi(\rho)\otimes |0\rangle\langle 0|)\\
   &=& \max_\rho \Tr f(\Phi(\rho))+d_\Phi(d_\Omega-1)f(0).
\end{eqnarray*}
The equality in $(*)$ follows from Example~\ref{ExIdentity}.
\qed

The first counterexample channel to the additivity conjecture for the $p$-R\'enyi entropy
(for $p>4.79$) has been given by Werner and Holevo~\cite{HolevoWerner}.
In dimension $d$, the Werner-Holevo channel $\Phi_d$ is defined as
\begin{equation}
  \Phi_d(\rho):=\frac 1 {d-1}\left(\idn-\rho^T\right).
  \label{HWChannel}
\end{equation}
It has the useful covariance property
\begin{equation}
   \Phi_d(U\rho U^\dagger)=\bar U\Phi_d(\rho)\bar U^\dagger
   \label{HWcovariance}
\end{equation}
for every unitary $U$. As a simple example, we derive a necessary condition
for additivity for this channel in dimension $d=3$:
\begin{example}
\label{ExHW}
If $f:[0,1]\to\R$ is a convex function with
\[
   f\left(\frac 1 3\right)+8 f\left(\frac 1 {12}\right)>5 f(0)+4 f\left(\frac 1 4\right),
\]
then $f$ is not additive for the Werner-Holevo channel pair $(\Phi_3,\Phi_3)$.
\end{example}
\proof
If $|\psi\rangle,|\varphi\rangle\in\C^d$ are arbitrary pure states, the output
$\Phi_d\otimes\Phi_d(|\psi\rangle\langle\psi|\otimes|\varphi\rangle\langle\varphi|)$ has a $(2d-1)$-fold
degenerate eigenvalue $0$, and a $(d-1)^2$-fold degenerate eigenvalue $1/(d-1)^2$. Due to the covariance property~(\ref{HWcovariance}),
this is true for all pure states and does not depend on $|\psi\rangle$ or $|\varphi\rangle$.
Thus,
\begin{eqnarray*}
   \Tr f(\Phi_d\otimes\Phi_d(|\psi\rangle\langle\psi|\otimes|\varphi\rangle\langle\varphi|))=\\
   (2d-1)f(0)+(d-1)^2 f\left(\frac 1 {(d-1)^2}\right).
\end{eqnarray*}
On the other hand, if we input a maximally entangled state $\rho_m$, it is shown in \cite{HolevoWerner}
that the output $\Phi_d\otimes\Phi_d(\rho_m)$ has a single eigenvalue $(2-2/d)/(d-1)^2$ and a $(d^2-1)$-fold
degenerate eigenvalue $(1-2/d)/(d-1)^2$, such that
\[
   \Tr f(\Phi_d^{\otimes 2}(\rho_m))=f\left(\frac{2-\frac 2 d}{(d-1)^2}\right)
   +(d^2-1)f\left(\frac{1-\frac 2 d}{(d-1)^2}\right).
\]
Comparing both expressions for $d=3$, we see that $f$ is not additive for $(\Phi_3,\Phi_3)$ if
the stated inequality holds.\qed

It is clear that we get more similar inequalities for $d\geq 4$, but in most cases,
these inequalities seem to be weaker.

The following lemma shows that the multiplicativity problem of the minimum output rank also fits
into Definition~\ref{DefEntFn}. We need this result later in the proof of Example~\ref{LemContinuous}.
\begin{lemma}[Minimum Output Rank]
\label{LemMinOutputRank}
\lineclear
The convex function
\[
   \delta_0(x):=\left\{
      \begin{array}{cl}
         1 & \mbox{ if }x=0\\
         0 & \mbox{ if }x\in(0,1]
      \end{array}
   \right.
\]
is not for all channels additive.
\end{lemma}
\proof
The function $\delta_0$ is related to the minimum output rank of quantum channels $\Phi$:
\[
   \max_\rho \Tr\, \delta_0(\Phi(\rho))=d-\min_\rho {\rm rank}(\Phi(\rho)),
\]
where $d$ is the dimension of the output Hilbert space of $\Phi$.
It has been shown in~\cite{MinRank} that the minimum output rank is not multiplicative;
there exist channels $\Phi$ and $\Omega$ such that
\[
   \min_\rho {\rm rank}(\Phi\otimes\Omega(\rho))<\min_\rho {\rm rank}(\Phi(\rho))\cdot \min_\rho {\rm rank}(\Omega(\rho)),
   \label{EqMinRank}
\]
which means that ${\rm rank}(\Phi\otimes\Omega(\rho))$ does not achieve its global minimum at tensor
product input states $\rho$. Consequently, $\Tr\,\delta_0(\Phi\otimes\Omega(\rho))$ achieves its global maximum at
entangled input states $\rho$.\qed

\section{On the Structure of Additive Functions}
Since every convex function on $[0,1]$ is bounded, the sup norm distance $\|f-g\|_\infty:=\sup_{x\in[0,1]} |f(x)-g(x)|$
can be used as a distance measure on the set of convex functions $\mathcal{F}$ on the unit interval. This way, we get a
notion of ``open'' and ``closed'' sets in $\mathcal{F}$. Formally, we get
the relative topology of $\mathcal{F}$ within the larger Banach space of bounded functions on $[0,1]$.

The next lemma shows that the set $M\subset\mathcal{F}$ of additive functions for a fixed pair of channels is a closed cone,
where ``cone'' refers to the simple property that $f\in M\Rightarrow \alpha f\in M$ holds for every $\alpha\geq 0$.
\begin{lemma}
\label{LemOpenCone}
With respect to the $\|\cdot\|_\infty$-norm topology, the set of additive functions on
a pair of channels $(\Phi,\Omega)$ is a closed cone.
\end{lemma}
\proof
The cone property is trivial: a function $f$ is additive for $(\Phi,\Omega)$ if and only if $\alpha\cdot f$ is
additive for $(\Phi,\Omega)$.

On the other hand, a function $f$ is {\em not} additive for $(\Phi,\Omega)$ if and only if there exists an entangled
state $\rho$, such that
\begin{equation}
   \Tr f(\Phi\otimes\Omega(\rho))>\Tr f(\Phi(\sigma_1)\otimes\Omega(\sigma_2))\quad \forall \sigma_1,\sigma_2.
   \label{EqNonAdditive}
\end{equation}
It is clear that there is some $\varepsilon>0$ such that for any convex function $g:[0,1]\to\R$ with $\|f-g\|_\infty<\varepsilon$,
equation~(\ref{EqNonAdditive}) still holds if $f$ is replaced by $g$. This shows that the set of non-additive functions
for $(\Phi,\Omega)$ is open.\qed

In the following, we will often show that a sequence of additive convex functions $\{f_n\}$ on $[0,1]$ converges pointwise
to a limit function $f$, and then refer to Lemma~\ref{LemOpenCone} to conclude that $f$ must be additive, too.
In fact, it is shown in \cite[Corollary 1.3.8]{Niculescu} that in this case, pointwise convergence implies
uniform convergence, and the limit function must be convex. Hence this kind of reasoning is justified.

It is a natural question whether the set of additive or non-additive functions has interesting properties.
One useful property is convexity. It is not clear in general if the set of additive functions
for a given arbitrary pair of channels is convex. However, convexity holds for the special class
of {\em unitarily covariant} channels. In accordance with \cite{FukudaUnitary}, we call a channel $\Phi$
unitarily covariant if for every unitary $U$, there exists a unitary $V$ such that
\begin{equation}
   \Phi(U\rho U^\dagger)=V\Phi(\rho)V^\dagger\qquad\mbox{for all }\rho.
   \label{EqUniCov}
\end{equation}
Sometimes a different class of channels is studied: a channel $\Phi$ is called {\em irreducibly covariant}
(cf.~\cite{DattaHolevoSuhov, Holevo})
if there are irreducible unitary representations $U_g,V_g$ of a group $G$ such that
\begin{equation}
   \Phi(U_g \rho U_g^\dagger)=V_g \Phi(\rho) V_g^\dagger
   \label{EqUniIrred}
\end{equation}
for all $g\in G$ and all $\rho$. Unitarily covariant channels need not be irreducibly covariant, and vice versa;
for example, if $\Phi$ and $\Omega$ are $d$-dimensional unitarily covariant channels with $V=U$, then
the tensor product channel $\Phi\otimes\Omega$ is irreducibly covariant with respect to $U(d)\times U(d)$,
but it is in general not unitarily covariant. For a counterexample in the opposite direction, define a channel
$\Omega$ on $\C^2$ via $\Omega(\rho):=(\Tr \rho)|0\rangle\langle 0|$, where $|0\rangle\in\C^2$ is some normalized vector.
Then $\Omega$ is unitarily covariant in the sense of Equation~(\ref{EqUniCov}) (with $V=\idn$ for every $U$),
but it is not irreducibly covariant, since any group representation $V_g$ satisfying Equation~(\ref{EqUniIrred})
must leave the subspace spanned by $|0\rangle$ invariant.

The Werner-Holevo channel (\ref{HWChannel}) is an example of a unitarily covariant channel due to (\ref{HWcovariance}).

\begin{lemma}[Unitarily Covariant Channels]
\label{LemConvexity}
\lineclear
If $\Phi$ and $\Omega$ are unitarily covariant channels, then the set
of additive functions on $(\Phi,\Omega)$ is convex (and due to Lemma~\ref{LemOpenCone}, a closed convex cone).
\end{lemma}
\proof Let $f$ and $g$ be additive convex functions for $(\Phi,\Omega)$. We have to prove that
$f+g$ is also additive for $(\Phi,\Omega)$.

Due to the unitary covariance of $\Phi$ and $\Omega$, the eigenvalues of $\Phi\otimes\Omega(\rho)$
and $\Phi\otimes\Omega(U\otimes V \rho U^\dagger \otimes V^\dagger)$ are the same for every unitary $U$ and $V$.
Thus, $\Tr f(\Phi\otimes\Omega(\rho))$ depends only on the Schmidt coefficients of the pure state $\rho$ (and similarly
for $g$). Since $f$ is additive, the expression $\Tr f(\Phi\otimes\Omega(\rho))$ attains its global
maximum at every pure unentangled input state $\rho$ at once. The same is true for $g$; thus, $f$ and
$g$ have a global maximizer in common. It follows that $f+g$ must have the same global maximizer, namely,
an unentangled state.\qed

The minimum output entropy additivity conjecture is known to hold true for the Werner-Holevo channel,
defined in (\ref{HWChannel}), in arbitrary dimensions.
According to Datta~\cite{Nilanjana} and Alicki and Fannes~\cite{AlickiFannes},
the same is true for the additivity of the $p$-R\'enyi entropy
for $1<p\leq 2$, but additivity does not hold if $p>4.79$ (cf. \cite{HolevoWerner}). Moreover, additivity also holds in
the domain $0<p<1$, as remarked in~\cite{MinRank}.

Due to Lemma~\ref{Lemma2},
those additivity results are related to the functions
$x\log x$ and $x^p$ for $1<p\leq 2$ as well as $-x^p$ for $0<p<1$. An interesting observation is
that all these functions are operator convex as defined in (\ref{EqOperatorConvex}). Thus, the following theorem
contains many known results on the Werner-Holevo channel as special cases:
\begin{theorem}[Werner-Holevo Channel]
\label{HWTheorem}
\lineclear
Every operator convex function $f:[0,\infty)\to\R$ is additive for
the Werner-Holevo channel (tensored with itself) in dimension $3$.
\end{theorem}
We conjecture that this is also true for the Werner-Holevo channel in larger dimensions $d\geq 4$
and for more than two factors; yet,
it seems that the original calculations
in \cite{Nilanjana} cannot be so easily adapted to that general case. Also, numerically
it seems that it is sufficient that $f$ is operator convex on $[0,1]$ (instead of $[0,\infty)$), but the proof is more difficult.

\proof It is well-known \cite{Bhatia,Nielsen} that every operator convex function $g$ on $(-1,1)$ has an integral representation
of the form
\[
   g(t)=g(0)+g'(0)t+\frac{g''(0)}2 \int_{-1}^1 \frac{t^2}{1-\lambda t} d\mu(\lambda),
\]
where $\mu$ is some probability measure on $[-1,1]$. Therefore, if $f$ is operator convex on $[0,\infty)$,
then it is in particular operator convex on $(0,1)$, and we can shift the above expression by substituting $x:=\frac{t+1}2$ to
obtain
\[
   f(x)=\alpha+\beta x +\gamma
   \int_{-1}^1 \frac{(2x-1)^2}{1-\lambda(2x-1)}d\mu(\lambda),
\]
where $\alpha+\beta x=f\left(\frac 1 2\right)+\frac 1 2 f'\left(\frac 1 2\right)(2x-1)$, and $\gamma=\frac 1 8 f''\left(\frac 1 2\right)\geq 0$.
Moreover, the measure $\mu$ must vanish on $(0,1]$, because $f_\lambda(x):=\frac{(2x-1)^2}{1-\lambda(2x-1)}$ has a pole
in the positive reals for every $\lambda\in(0,1]$, but $f$ is by assumption defined on all of $[0,\infty)$. For the same
reason, $\mu$ must vanish at $\lambda=-1$.

According to Lemma~\ref{LemConvexity}, it is thus sufficient to
show that the functions $\alpha+\beta x$ and $f_\lambda$ are additive for every $\lambda\in(-1,0]$; then,
it follows that $f$ must be additive, too.

But the function $\alpha+\beta x$ is trivially additive (as shown in Lemma~\ref{LemAffine}). Let $\Phi_3$
be the Werner-Holevo channel in dimension $d=3$ as defined in (\ref{HWChannel}).
From \cite{Datta}, we know the eigenvalues of the output $\Phi_3\otimes\Phi_3(\rho)$ if the
input has Schmidt coefficients $(\lambda_1,\lambda_2,\lambda_3)$: There are $6$ eigenvalues of the form
\[
   e_{\alpha\beta}:=\frac {1-\lambda_\alpha-\lambda_\beta}4\qquad(\alpha\neq \beta,\quad\alpha,\beta=1,2,3)
\]
and $3$ eigenvalues of the form
\[
   G_\alpha:=\frac 1 3 \cos^2\left(\frac\theta 6 -\frac{2 \pi (\alpha-1)}6\right)\qquad(\alpha=1,2,3),
\]
where $\tan\theta=\frac{\sqrt{t\left(\frac 1 {27}-t\right)}}{t-\frac 1 {54}}$, and $t=\lambda_1\lambda_2\lambda_3$. Hence,
\[
   \Tr f_\lambda(\Phi_3\otimes\Phi_3(\rho))=\sum_{\alpha\neq\beta} f_\lambda(e_{\alpha\beta})+\sum_{\alpha=1}^3 f_\lambda(G_\alpha).
\]
Since the set
\[
   \left\{\left.\left(\frac{1-\lambda_\alpha-\lambda_\beta}4\right)_{\alpha\neq\beta}\,\,\right|\,\, \sum_{i=1}^3 \lambda_i=1\right\}\subset \R^6
\]
is convex, the convex function $\sum_{\alpha\neq\beta} f_\lambda(e_{\alpha\beta})$ attains its maximum on the
extremal points, i.e. those points where, up to permutation, $\lambda_1=1$ and $\lambda_2=\lambda_3=0$.

Due to the simple form of the functions $f_\lambda$, it is easy to show with some analysis that the function
$\sum_{\alpha=1}^3 f_\lambda(G_\alpha)$ attains its global maximum for $\theta\in[0,\pi]$ at
$\theta=\pi$, corresponding to $t=\lambda_1\lambda_2\lambda_3=0$. In fact, this expression is constant in $\theta$ for $\lambda=0$, and
it is increasing in $\theta$ if $-1<\lambda<0$.

In summary, $\Tr f_\lambda(\Phi_3\otimes\Phi_3(\rho))$ attains its global maximum on the states with Schmidt
coefficients $(1,0,0)$, i.e. on the unentangled states. Thus, $f_\lambda$ is additive for every $\lambda\in(-1,0]$
for two copies of the Werner-Holevo channel in dimension $3$.
The claim follows.\qed

Consider the set $\mathcal{U}$ of functions that are additive for {\em all} pairs of unitarily covariant
channels $(\Phi,\Omega)$. According to Lemma~\ref{LemConvexity}, the set $\mathcal{U}$ is a closed convex
cone. It is an interesting problem to determine the set $\mathcal{U}$ explicitly.
In the light of Theorem~\ref{HWTheorem}, and due to the fact that the most natural closed convex subset
of the convex functions is the set of operator convex functions, the following conjecture seems natural:
\begin{conjecture}[Additivity\&Operator Convexity]
\label{ConjAddOp}
The set of functions $\mathcal{U}$ that are additive for all unitarily covariant channels
agrees with the set of operator convex functions on some interval $I\subset\R$.
\end{conjecture}
It seems that for a {\em fixed} pair of channels, the set of additive functions does not
have a simple description in general, and several natural conjectures on the structure of the set of additive functions
fail. For example, it is easy to construct convex functions $f$ and $g$ such that $f$ and $f+g$ are additive
for the Werner-Holevo channel pair $(\Phi_3,\Phi_3)$, but such that $g$ is not additive for $(\Phi_3,\Phi_3)$.
Also, there are additive functions $f$ and $g$ such that $\max\{f,g\}$ is not additive (cf. Theorem~\ref{ThePiecewiseAffine}).

In the following, we will prove some more results on the set of functions that are additive for certain sets of channels.
We will assume that the channel sets have the following property:
\begin{definition}[Channel Classes]
\label{DefChannelClass}
In the remainder of the paper, a {\em channel class} $\mathcal{C}$ is a set of channels which is closed
with respect to tensor products, and which contains all maximally depolarizing channels. That is,
\begin{itemize}
\item $\Phi,\Omega\in\mathcal{C}\Rightarrow \Phi\otimes\Omega\in\mathcal{C}$,
\item $\Sigma_\sigma\in\mathcal{C}$ for all $\sigma=\frac 1 d\idn$, where $\Sigma_\sigma(A):=\Tr(A)\sigma$.
\end{itemize}
\end{definition}
Examples of channel classes are
\begin{itemize}
\item the set of all channels, and
\item the set of irreducibly covariant channels: tensor products of irreducibly covariant
channels are again irreducibly covariant~\cite{DattaHolevoSuhov}, and $\Sigma_\sigma$ is irreducibly
covariant if $\sigma$ is proportional to the identity.
\end{itemize}
The set of unitarily covariant channels is not a channel class. However, the set of all channels which can be written
as tensor products of unitarily covariant channels is a channel class.

We are interested in the set of functions that are additive for all channels in a given channel class $\mathcal{C}$
(we call them the ``functions that are additive for $\mathcal{C}$''). According to Lemma~\ref{LemOpenCone},
the set of those functions is a closed cone for every channel class $\mathcal{C}$. But we can say more.
\begin{theorem}
\label{TheTensorStructure}
If a convex function $f:[0,1]\to\R$ is additive for a channel class $\mathcal{C}$ (as defined in Definition~\ref{DefChannelClass}),
then $f\left(\frac x n\right)$
is additive for $\mathcal{C}$ for every $n\in\N$, too.

Moreover, if $\mathcal{C}$ is additionally closed with respect to tensor products with $\Sigma_\sigma$ for all $\sigma$,
then the function
\[
   x\in[0,1]\mapsto \sum_{i=1}^n f(\mu_i\cdot x)
\]
is additive for $\mathcal{C}$ as well for every probability vector $(\mu_1,\ldots,\mu_n$).
\end{theorem}
\proof Let $\vec \mu$ be an arbitrary probability vector, and let $f$ be additive for $\mathcal{C}$. Let $\Phi,\Omega\in\mathcal{C}$
be arbitrary channels. We have to show that the function $\tilde f(x):=\sum_{i=1}^n f(\mu_i x)$ is
additive for $(\Phi,\Omega)$.

Let $\sigma$ be a $n\times n$ density operator with eigenvalues $\mu_1,\ldots,\mu_n$. Consider the channel $\Phi\otimes\Omega\otimes\Sigma_\sigma$,
and let $|\psi\rangle$ be an input vector for this tripartite channel. It has a Schmidt decomposition
\[
   |\psi\rangle=\sum_i \sqrt{\lambda_i} |i_{\Phi\Omega}\rangle\otimes|i_{\Sigma_\sigma}\rangle,
\]
where $\{|i_{\Phi\Omega}\rangle\}_i$ and $\{|i_{\Sigma_\sigma}\rangle\}_i$ are orthonormal bases on the input Hilbert spaces
for the channels $\Phi\otimes\Omega$ and $\Sigma_\sigma$ respectively. The corresponding output is
\begin{eqnarray*}
   \Phi\otimes\Omega&\otimes&\Sigma_\sigma\enspace(|\psi\rangle\langle\psi|)\\
   &=&\sum_{ij}\sqrt{\lambda_i\lambda_j}
   \Phi\otimes\Omega(|i_{\Phi\Omega}\rangle\langle j_{\Phi\Omega}|)\otimes\Sigma(|i_{\Sigma_\sigma}\rangle\langle j_{\Sigma_\sigma}|)\\
   &=&\sum_i \lambda_i \Phi\otimes\Omega(|i_{\Phi\Omega}\rangle\langle i_{\Phi\Omega}|)\otimes\sigma.
\end{eqnarray*}
The trace of a convex function on that output attains its maximum, due to convexity, in the extremal case
where, up to permutation, $\lambda_1=1$ and $\lambda_2=\lambda_3=\ldots=0$. This means that we may choose
the input to be unentangled between $\Phi\otimes\Omega$ and $\Sigma_\sigma$. In this case, if the output
$\Phi\otimes\Omega(|\psi\rangle\langle\psi|)$ has spectrum $\{\alpha_1,\ldots,\alpha_N\}$, then the
output $\Phi\otimes\Omega\otimes\Sigma_\sigma(|\psi\rangle\langle\psi|\otimes|\varphi\rangle\langle\varphi|)$ has
spectrum $\{\alpha_i\mu_j\}_{i,j}$.
Since $f$ is additive for $\mathcal{C}$, it is in particular additive for the channel pair  $(\Phi\otimes\Sigma_\sigma,\Omega)$
as long as $\mathcal{C}$ is closed with respect to tensor products with $\Sigma_\sigma$.
If this is the case,
the expression $\Tr f(\Phi\otimes\Omega\otimes\Sigma_\sigma(|\psi\rangle\langle\psi|\otimes|\varphi\rangle\langle\varphi|)$
attains its global maximum at an unentangled input state $|\psi\rangle$. But
\begin{eqnarray*}
\Tr \tilde f(\Phi\otimes\Omega(|\psi\rangle\langle\psi|))&=&\sum_i \tilde f(\alpha_i)
=\sum_i \sum_j f(\mu_j \alpha_i)\\
&=&\Tr f(\Phi\otimes\Omega\otimes\Sigma_\sigma(|\psi\rangle\langle\psi|\otimes|\varphi\rangle\langle\varphi|)),
\end{eqnarray*}
and so the expression $\Tr \tilde f(\Phi\otimes\Omega(|\psi\rangle\langle\psi|))$ attains its global maximum
at unentangled input states $|\psi\rangle$. It follows that $\tilde f$ is additive for $(\Phi,\Omega)$.
In particular, we get that $f\left(\frac x n\right)$ is additive for $\mathcal{C}$ for every $n\in\N$ if we insert
$\vec \mu=\left(\frac 1 n,\frac 1 n,\ldots,\frac 1 n \right)$.
\qed

As a simple example application, we find that functions which are additive for {\em all} channels
must be continuous at zero:
\begin{example}
\label{LemContinuous}
If a convex function $f:[0,1]\to\R$ is additive for all channels, then it is continuous at zero.
\end{example}
\proof
Let $\mathcal{C}$ be the class of all channels.
Suppose that $f$ is additive for $\mathcal{C}$, but {\em not} continuous at zero. Since $f$ is convex, the limit
$y:=\lim_{x\to 0} f(x)$
exists and is less than $f(0)$. If $f$ is additive for $\mathcal{C}$, then the function
\[
   g(x):=\frac{f(x)-y}{f(0)-y}
\]
is additive for $\mathcal{C}$ as well due to Example~\ref{LemAffine}. Since $f$ is continuous on $(0,1)$ and
\[
   g\left(\frac x n\right)=\left\{
      \begin{array}{cl}
         1 & \mbox{if }x=0\\
         \frac{f\left(\frac x n\right)-y}{f(0)-y} & \mbox{if }x\in(0,1],
      \end{array}
   \right.
\]
the sequence of functions $\left\{g\left(\frac x n\right)\right\}_{n\in\N}$ converges
to the function $\delta_0$ introduced in Lemma~\ref{LemMinOutputRank}. But we know from Theorem~\ref{TheTensorStructure}
that the functions $g\left(\frac x n\right)$ are additive for $\mathcal{C}$ for every $n\in\N$. Moreover, according to
Lemma~\ref{LemOpenCone}, the set of additive functions for $\mathcal{C}$ is closed. Thus, $\delta_0$ must be additive,
which contradicts Lemma~\ref{LemMinOutputRank}.\qed

We will later see that this result is not valid for $x=1$: in Lemma~\ref{LemDelta1},
we show that there exist additive functions that are discontinuous at $x=1$.

Here is another interesting example which in some sense ``interpolates'' between the von Neumann and $p$-R\'enyi entropies:
\begin{example}[Distorted Entropy and $p$-Purity]
\label{LemDistortedEntropy}
\lineclear
Let $\frac 1 2 \leq p \leq 1$ and $\mathcal{C}$ a channel class. If the function
\[
   x^p \log x
\]
is additive for $\mathcal{C}$, then the function $-x^p$ is additive for $\mathcal{C}$ as well; consequently, the minimum output
$p$-R\'enyi entropy is additive for all channel pairs in $\mathcal{C}$.
\end{example}
\proof Notice that $x^p\log x$ is convex on $[0,1]$ if and only if $\frac 1 2 \leq p \leq 1$,
which explains the choice of the interval for $p$.
Suppose that $x^p\log x$ is additive for $\mathcal{C}$. Then, $\left(\frac x n\right)^p \log \frac x n$ is additive for $\mathcal{C}$
as well for every $n\in\N$ according to Theorem~\ref{TheTensorStructure}. As multiplication with a constant
does not affect additivity, it follows that $x^p\log x - x^p \log n$ is additive for $\mathcal{C}$ as well, and so is
\[
   -x^p+\frac{x^p\log x}{\log n}\qquad\mbox{for every }n\in\N.
\]
Taking the limit $n\to\infty$, the claim follows from Lemma~\ref{LemOpenCone}.\qed

Here are some more consequences of Theorem~\ref{TheTensorStructure}. The proofs are very similar
to the proof of Example~\ref{LemDistortedEntropy} and thus omitted.
\begin{lemma}[von Neumann Entropy, Analyticity]
\label{LemVonNeumann}
Let $f:[0,1]\to\R$ be a convex function.
\begin{itemize}
\item If $f(x)=ax\log x +\mathcal{O}(x)$ with $a\neq 0$ and $f$ is additive for a class of channels $\mathcal{C}$,
then $x\log x$ is additive for $\mathcal{C}$ as well, i.e. the minimum
output von Neumann entropy is additive for $\mathcal{C}$.
\item If $f$ has a non-linear analytic extension to a complex neighbourhood of zero, then there exist channels $(\Phi,\Omega)$
such that $f$ is not additive for $(\Phi,\Omega)$.
\end{itemize}
\end{lemma}
This shows that von Neumann entropy plays some kind of special role: if any function that behaves like $x\log x$
for small $x$ is additive, then von Neumann entropy is automatically additive as well. The second part of the lemma
concerns possible functions $f$ that are additive for {\em all} channels: this possibility is ruled out for many
functions, for example, say, for $f(x)=\frac 1 {1+ax}$ for $a>-1$. The main idea to prove the second part
is the fact that, after subtracting a linear function, analytic functions can be approximated by a monomial $a\cdot x^m$,
but the functions $x^m$ violate additivity for some channels as shown, for example, in~\cite{WinterComm}.

A convex function on $[0,1]$ is automatically continuous on $(0,1)$, but it may be
discontinuous at the endpoints. We have shown in Example~\ref{LemContinuous} that functions that
are additive for all channels are continuous at $x=0$. In contrast, the following simple arguments show that
additive functions may be discontinuous at $x=1$.
\begin{lemma}
\label{LemDelta1}
If $\Phi$ and $\Omega$ are quantum channels such that $\Phi\otimes\Omega$ outputs a pure state,
then every convex function is additive for $(\Phi,\Omega)$.

Consequently, if $f,g:[0,1]\to\R$ are
convex functions that differ only at $x=1$, then $f$ is additive for any pair of channels if and only if
$g$ is additive for that pair of channels.
\end{lemma}
\proof
Let $f:[0,1]\to\R$ be a convex function, and suppose there exists some input state $\rho_0$
and a pure state $|\varphi\rangle$
such that $\Phi\otimes\Omega(\rho_0)=|\varphi\rangle\langle\varphi|$.
Denoting the minimum output entropy of a channel $\Phi$ by $S_{min}(\Phi)$ as in the introduction,
it is well-known \cite{Hayden} and in fact proven in the present paper in Lemma~\ref{LemSingleChannel} that
\[
   0=S_{min}(\Phi\otimes\Omega)\geq S_{min}(\Phi),
\]
and so $\Phi$ (and by the same argument, $\Omega$) outputs a pure state, too.
Taking the tensor product of the corresponding inputs, we get an unentangled (pure tensor product)
input state $\tilde\rho_0$ for $\Phi\otimes\Omega$ such that $\Phi\otimes\Omega(\tilde\rho_0)$ is
pure as well. Due to the Schur convexity~\cite{Bhatia} of the map $(\lambda_1,\ldots,\lambda_n)\mapsto
\sum_{i=1}^n f(\lambda_i)$, the state $\tilde\rho_0$ is a maximizer of the map $\rho\mapsto\Tr f(\Phi\otimes\Omega(\rho))$,
and so $f$ is additive for $(\Phi,\Omega)$.

Let now $\Phi$ and $\Omega$ be arbitrary quantum channels.
If $\Phi\otimes\Omega$ outputs a pure state, then both $f$ and $g$ are additive for $(\Phi,\Omega)$.
On the other hand, if $\Phi\otimes\Omega$ does not output a pure state, then
the eigenvalues of every output are strictly less than $1$, and $\Tr f(\sigma)=\Tr g(\sigma)$
for every $\sigma$.\qed

Thus, modifying a convex function at $x=1$ does not affect its additivity property.
For example, the following function $\delta_1$ is additive for all channels:
\[
   \delta_1(x):=\left\{
      \begin{array}{cl}
         0 & \mbox{ if }x\in[0,1)\\
         1 & \mbox{ if }x=1.
      \end{array}
   \right.
   \label{EqDelta1}
\]

\section{Piecewise Linear Functions}
The simplest functions that have not yet been studied before in the context of additivity
are the piecewise linear functions. More in detail, while linear functions $f(x)=ax+b$
are additive for all channels according to Lemma~\ref{LemAffine}, the simplest examples of functions
with unknown additivity properties are those functions $f$ which are the maximum of two
linear functions,
\begin{equation}
   f(x):=\left\{
      \begin{array}{cl}
         ax+b & \mbox{if }x\leq x_0\\
         cx+d & \mbox{if }x>x_0
      \end{array}
   \right.
   \label{EqPiecewise}
\end{equation}
with $a x_0+b=c x_0+d$, and $a<c$ to ensure continuity and convexity.
We call $x_0$ the {\em kink} of $f$.

\begin{figure}[!hbt]
\includegraphics[angle=0, width=4cm]{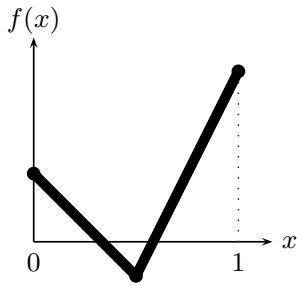}
\caption{A piecewise linear function.}
\label{AbbKink}
\end{figure}

Fig.~\ref{AbbKink} shows what such functions look like.
Are those functions additive? In this section, we give a partial answer to
this question. We first note a simple consequence of Theorem~\ref{TheTensorStructure}:
there we have shown that if $f$ is additive, then $f\left(\frac x n\right)$ must be additive as well.
It is natural to conjecture that more generally, $f(\lambda x)$ must then always be additive
for every $\lambda\in[0,1]$. While it is not clear if this holds true in general, we can prove it
for the case that $f$ is differentiable at zero:
\begin{lemma}
\label{LemLambda}
Let $f:[0,1]\to\R$ be a convex function which is differentiable at zero. If $\mathcal{C}$ is a class
of channels which is closed with respect to tensor products with $\Sigma_\sigma$ for all $\sigma$
(cf. Definition~\ref{DefChannelClass}), and if $f$ is additive for $\mathcal{C}$, then the function
\[
   x\in[0,1]\mapsto \sum_{i=1}^n f(\mu_i\cdot x)
\]
is additive for $\mathcal{C}$ as well for every sub-probability vector $(\mu_1,\ldots,\mu_n)$, i.e.
if $\mu_i\geq 0$ for every $1\leq i \leq n$ and $\sum_{i=1}^n \mu_i\leq 1$.

In particular, $f(\lambda x)$ is additive for $\mathcal{C}$ for all $\lambda\in[0,1]$.
\end{lemma}
\proof
Without loss of generality, we may assume that $f(0)=0$, otherwise we can add
some constant to $f$ without changing its additivity properties.

Let $m:=1-\sum_{i=1}^n \mu_i$, then $\left(\underbrace{\frac m N,\ldots,\frac m N}_N,\mu_1,\ldots,\mu_n\right)$
is a probability vector for every $N\in\N$. According to Theorem~\ref{TheTensorStructure}, the function
\[
   N\cdot f\left(\frac m N x\right)+\sum_{i=1}^n f(\mu_i x)
\]
must then be additive for $\mathcal{C}$ for every $N\in\N$. Since $f$ is by assumption
differentiable at zero, the limit $\lim_{h\to 0} \frac{f(h)} h$ exists and equals $f'(0)$. Hence
\[
   \lim_{N\to\infty} N\cdot f\left(\frac {mx}N\right)=mx\lim_{N\to\infty}\frac N {mx} f\left(\frac{mx}N\right)=mxf'(0).
\]
Due to the closedness property of the additive functions as shown in Lemma~\ref{LemOpenCone},
it follows that the function
\[
   \sum_{i=1}^n f(\mu_i x)+m f'(0) x
\]
is additive for $\mathcal{C}$. But $m f'(0)x$ is a linear function that we may subtract without affecting additivity
due to Example~\ref{LemAffine}.\qed

We now use this lemma to prove our result on piecewise linear functions: if those functions
are additive or not depends only on the location of the kink.
\begin{theorem}[Piecewise Linear Functions]
\label{ThePiecewiseAffine}
\lineclear
There is a global constant $\frac 1 3\leq\gamma\leq 1$ such that the following holds true:
if $f$ is the maximum of two linear functions as plotted in Fig.~\ref{AbbKink}, with kink at $x_0$,
then
\[
   f\mbox{ is additive for all channels }\Leftrightarrow x_0\geq \gamma.
\]
Similarly, for every channel class $\mathcal{C}$ which is closed with respect to tensor products with $\Sigma_\sigma$
for all $\sigma$, there is a constant $0\leq\gamma_{\mathcal{C}}\leq 1$ with the same property.
\end{theorem}
It is natural to conjecture that $\gamma=1$ holds; in this case,
no function of this type would be additive for all channels.

\proof
For simplicity, we assume that $\mathcal{C}$ is the class of all channels; the more general case
is completely analogous.
It is sufficient to consider the piecewise linear functions
\[
   g_{x_0}(x):=\left\{
      \begin{array}{cl}
         0 & \mbox{if }x\leq x_0\\
         x-x_0 & \mbox{if }x>x_0
      \end{array}
   \right.
\]
since every function which is the maximum of two linear functions can be transformed into
this form without affecting its additivity properties, if it has kink at $x_0$.
Explicitly, if $f$ is defined as in (\ref{EqPiecewise}), then the function $g$ defined by
\[
   g(x):=\frac{f(x)-(ax+b)}{c-a}
\]
has this form, and shares the additivity property with $f$ due to Example~\ref{LemAffine}.

Thus, additivity of $f$ (resp. $g$) depends only on the location of the kink. Now suppose
$g_t$ is additive for some $t\in[0,1]$. As $g_t$ is differentiable at zero, it follows from
Lemma~\ref{LemLambda} that $g_t(\lambda x)$ is additive as well for every $\lambda\in(0,1)$.
It is elementary to see that
\[
   \frac 1 \lambda g_t(\lambda x)=g_{\frac t \lambda}(x),
\]
and so $g_{\frac t \lambda}$ is additive, or equivalently $g_{t'}$ for every $t'\geq t$.
This shows that there is some constant $\gamma\in[0,1]$ such that $g_{x_0}$ is additive
if and only if the kink $x_0$ is larger than or equal to $\gamma$.

Finally, if $\frac 1 4<x_0<\frac 1 3$, then $g_{x_0}$ is not additive according to Example~\ref{ExHW}.
This shows that $\gamma\geq \frac 1 3$.\qed

Here is a recipe how to improve the lower bound on $\gamma$ (or in the best case to
prove that $\gamma=1$): find an example of a pair of channels such that
the maximum output eigenvalue $\Lambda$ is attained at an entangled input state.
Then $\gamma\geq\Lambda$. In fact, the proof above (or rather its reference to Example~\ref{ExHW})
exploits this fact for a pair of Werner-Holevo channels in dimension $3\times 3$.

\section{Conclusions}
In this paper, we have studied the problem
whether a given convex trace function, if it is applied to the output of a bipartite quantum channel, attains its
maximum at an unentangled input state. This problem generalizes the minimum output entropy additivity problem
in a natural way: for example, there is a single channel bound on the output capacity
(Lemma~\ref{LemSingleChannel}), additivity always holds if one of the
channels is the identity channel (Example~\ref{ExIdentity}), and the study of the minimum output rank
(Lemma~\ref{LemMinOutputRank}) and the largest output eigenvalue (Theorem~\ref{ThePiecewiseAffine}) have
natural interpretations in our more general framework.

In Theorem~\ref{HWTheorem}, we have shown that all operator convex functions on $[0,\infty)$
are additive for the Werner-Holevo channel in $3\times 3$ dimensions, which contains the well-known
additivity results for this channel as special cases.
Since the set of functions that are
additive for all unitarily covariant channels is convex (Lemma~\ref{LemConvexity}), it is natural to conjecture
that this set of functions can be classified further, possibly in a way as stated in Conjecture~\ref{ConjAddOp}.

We have also shown some additional structural properties of the set of additive functions (e.g. Lemma~\ref{LemOpenCone}
or Theorem~\ref{TheTensorStructure}), drawing new connections between functions like $x^p\log x$ and the $p$-R\'enyi entropies,
and also yielding partial reasons why von Neumann entropy seems to play a special role for additivity (cf. Lemma~\ref{LemVonNeumann}).

Even though the original additivity conjecture has recently been disproved~\cite{Hastings}, it is still interesting
to study additivity for special classes of channels. Moreover, the transition from additivity to non-additivity
(say, the dimensionality of the channels) is still not well understood, and the history of the additivity
problem shows that introducing new entropy notions (like $p$-R\'enyi entropy) can be useful.
This is why we are confident that our framework of additive convex functions might be helpful in some instances of this problem.

\vskip 0.25cm
\noindent{\bf Acknowledgments.}
The author would like to thank N. Ay, J. Eisert, D. Gross, T. Kr\"uger, R. Seiler, A. Szko\l a,
R. Werner, and C. Witte
for helpful discussions -- and \-especially Ra.~Siegmund-Schultze for his never-ending enthusiasm for the additivity conjecture.

Special thanks go to A. Winter for his kind hospitality during a visit to Bristol and for many discussions.

\end{document}